\documentclass{amsart}

\usepackage{amsmath,amssymb,graphicx,hyperref,multirow}

\newcommand{\R}{\mathbb R}

\newcommand{\rar}{\rightarrow}

\DeclareMathOperator{\diag}{diag}
\DeclareMathOperator{\midd}{mid}
\DeclareMathOperator{\rank}{rank}

\begin{document}

\title[Formal Verification of Octorotor using Barrier Functions and SMT]{Formal Verification of Octorotor Flight Envelope using Barrier
  Functions and SMT Solving}
\address{HRL Laboratories, LLC\\
  Malibu, California}

\author{Byron~Heersink, Pape~Sylla, and~Michael~A.~Warren}
\maketitle

\begin{abstract}
This paper introduces an approach for formally verifying the safety of
the flight controller of an octorotor platform. Our method involves finding
regions of the octorotor's state space that are considered safe, and
which can be proven to be invariant with respect to the
dynamics. Specifically, exponential barrier functions are used to
construct candidate invariant regions near desired commanded
states. The proof that these regions are invariant is discovered
automatically using the dReal SMT solver, which ensures the accurate command
tracking of the octorotor to within a certain margin of error. Rotor
failures in which rotor thrusts become stuck at fixed values are
considered and accounted for via a pseudo-inverse control
allocator. The safety of the control allocator is verified in dReal
by checking that the thrusts demanded by the allocator never exceed
the capability of the rotors. We apply our approach on a specific
octorotor example and verify the desired command tracking properties
of the controller under normal conditions and various combinations of
rotor failures.
\end{abstract}

\section{Introduction}\label{Sec:intro}

Recently, interest in the study of multirotor air vehicles has been
growing quickly due to their high maneuverability and many
applications, such as inspection and surveillance. In particular, the
quadrotor has been a popular platform with which to conduct UAV
(unmanned aerial vehicle) research. While the quadrotor can
tolerate partial rotor faults (see, e.g., \cite{SMGZ,ZZRT,RK,ZC}), it
becomes uncontrollable if one of its rotors fails completely (though
\cite{FLL} shows that only yaw control need be lost). Thus the
octorotor has been studied as an alternative that is more robust to
rotor failures \cite{MWY,AE,AE2,SLFFSS,SLFSF}.

An approach to addressing flight safety for UAVs in a number of recent
works has been through the use of control barrier functions (CBFs). A
CBF is a scalar-valued function on the state space of a control system
whose support (here understood as the region where the function
attains positive values) is forward invariant under appropriate controls. Thus,
one can (mathematically) guarantee the safety of the system by finding
a CBF with support contained in a safe region of the state space. In
particular, in \cite{WAE}, CBFs are used to ensure that teams of
quadrotors are able to fly in a collision free manner; in \cite{NA},
CBFs are used for obstacle avoidance in quadrotor path planning in a
surveillance scenario; and in \cite{XS}, CBFs are used for flight
safety for quadrotors under some degree of human control. Other works
using barrier functions for UAV safety include
\cite{WS,WTE,KZC}. Other applications of barrier functions include 
adaptive cruise control and lane keeping \cite{AXGT,XWPGETGA,XGTA},
bipedal walking robots \cite{NS2,NHGAS}, and collision avoidance for
multirobot systems on land \cite{WAE2}. See \cite{ACENST} for a survey
of control barrier functions.

Instead of utilizing barrier functions within control algorithms to ensure
safety, as in the aforementioned references, the focus of this paper
is to use barrier functions to analyze the safety properties of
conventionally defined controllers. Specifically, our aim is to
provide a method of verifying the safety of controllers using
satisfiability modulo theories (SMT) solving. Our work is in the
spirit of \cite{JV2018,JV2019}, in which controllers for flight
vehicles are designed, and the flight envelope and asymptotic
stability of the controller are verified by the automated theorem
prover MetiTarski. However, in these works, the design of the
controllers is fairly specialized, and is closely tied together with
how the controller's safety is verified.  As above, a principal goal
of our work is to decouple safety analysis from controller design so
that our techniques can be employed even in cases where the underlying
controller can no longer be modified or where it is for other reasons
not feasible to incorporate barrier functions at the design stage.

The goal of this paper is therefore to introduce another approach to the formal
verification of the flight envelope of a UAV, one particularly that
has the potential to be used in a way that is independent of the
design of the controller. Our core method involves the use of barrier
functions, more specifically exponential barrier functions \cite{NS}
of a certain form, to produce an invariant subset of a prescribed safe
region near a desired commanded point in the control system's state
space (e.g., as one might see in connection with gain scheduling).
Our method aims to analyze systems with a controller designed to track
commands and whose dynamics can be approximated reasonably well by the
linearization around a desired operating state. In particular, our
approach lends itself well to analyzing UAV safety in states where the
vehicle is nearly upright or mildly tilting. It also has potential for
use in the verification of gain scheduled controllers that are
currently in common use for many kinds of aircraft.

The specific type of UAV we analyze is an octorotor, which, as
mentioned above, is robust to rotor failures. This allows us to
address the issue of fault tolerance in our approach. We account for
complete rotor failures, in which one or more rotors
stop working entirely and exert zero thrust, and more generally
failures in which rotors become stuck at specific speeds. To mitigate
these failures, we utilize a simple control allocator to attempt to
maintain the ideal octorotor behavior. The main property of the
control allocator that needs to be verified is that the rotors are
always capable of exerting the rotor thrusts commanded by the
allocator. We assume that failures are perfectly known, and do not
address the issue of fault detection, as in \cite{SLFFSS,SLFSF}. 

In Section \ref{Sec:model}, we present the equations of motion of the
octorotor model we study. In Section \ref{Sec:ctrl}, we describe the
controller. This includes an explanation of the kinds of rotor
failures considered and how they are accounted for by the control
allocator. Section \ref{Sec:barrier} introduces the general barrier
function framework employed, and Section \ref{Sec:octo_barrier}
describes precisely the form of barrier functions used for the
octorotor. In Section \ref{Sec:ver}, we formulate in detail the
properties that are analyzed in dReal and that confirm the desired
safety properties of the octorotor controller. In Section
\ref{Sec:results}, we describe the results obtained for a specific
controller: we formally verify that the octorotor will follow commands
closely in spite of dynamic disturbances when functioning normally and
under various rotor failures, and we identify some combinations of
rotor failures in which the desired conditions are violated. Finally,
concluding remarks are found in Section \ref{Sec:con}.

\section{Octorotor dynamics model}\label{Sec:model}

\begin{figure}
\centering
\includegraphics[width=0.7\columnwidth]{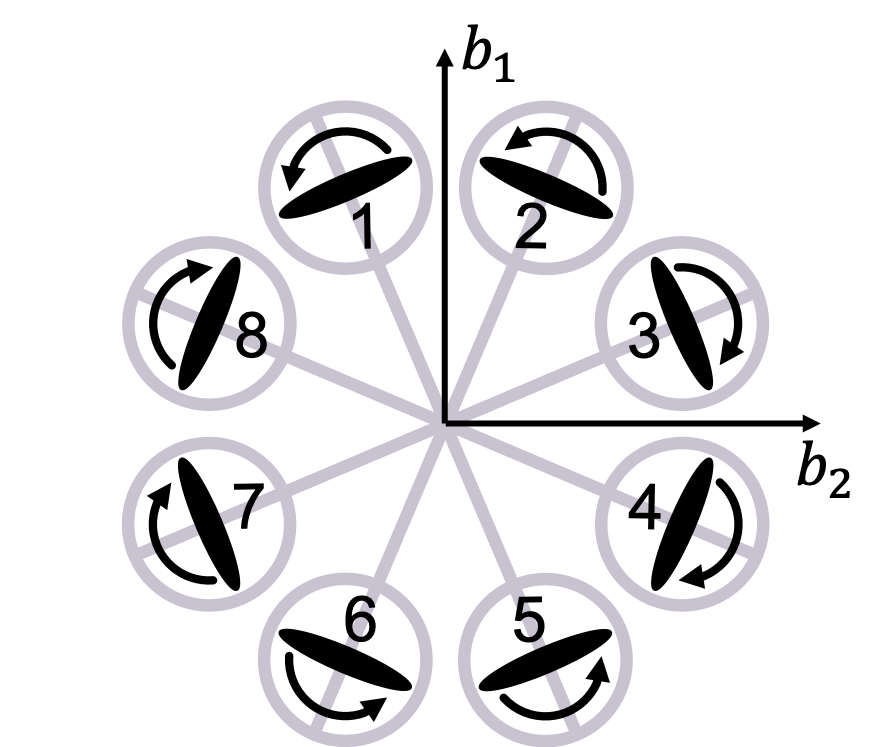}
\caption{Octorotor diagram}
\label{Fig:octo_diag}
\end{figure}

In this section, we describe the octorotor dynamics model that we
analyze. We follow \cite{GLL,LLM} in modeling the rigid body dynamics
and some characteristics of the rotors, and our octorotor model is
based on that in \cite{MWY}. See also \cite{MKC} for an introduction
to the modeling of multirotor aerial vehicles. The octorotor consists
of eight identical rotors arranged in the shape of a regular octagon as
illustrated in Figure \ref{Fig:octo_diag}. Each rotor exerts an upward
thrust normal to the plane of the octorotor, and a torque about the
octorotor's center of mass as explained below. 

To express the octorotor's dynamics, we first describe the two
reference frames required. First, the world reference
frame, which we assume is inertial, is given by the standard unit
basis vectors $e_1=[1,0,0]^T$, $e_2=[0,1,0]^T$, and
$e_3=[0,0,1]^T$. We assume that $e_1$ points north, $e_2$ points east,
and $e_3$ points down. Next, we assume that the body of the octorotor
has reference frame given by the orthonormal (vertical) vectors
$b_1,b_2,b_3$, where $b_1$ is considered the forward direction with
respect to the octorotor, $b_2$ is the right direction, and $b_3$ is
the down direction. (In Figure \ref{Fig:octo_diag}, $b_3$ points into
the page.) The $3\times3$ matrix $[b_1,b_2,b_3]$ is denoted by $R$,
which is the rotation matrix from the body frame to the world frame. 

Now we describe the state space of the system. We denote the
inertial position of the octorotor's center of mass by $r=[x,y,z]^T$;
and the inertial velocity by $v=[v_x,v_y,v_z]^T$.  Another component
of the state of the octorotor is its orientation, which is given by
$R$. Alternatively, we express the orientation in terms of Euler
angles. Letting $\phi,\theta,\psi$ be the roll, pitch, and yaw angles,
respectively, corresponding to the $Z$-$Y$-$X$ sequence for Euler
angles, we have the following relationship between $R$ and
$[\phi,\theta,\psi]$:
\begin{equation*}
\begin{aligned}
R&=
\begin{bmatrix}
c_\psi&-s_\psi&0\\
s_\psi&c_\psi&0\\
0&0&1
\end{bmatrix}
\begin{bmatrix}
c_\theta&0&s_\theta\\
0&1&0\\
-s_\theta&0&c_\theta
\end{bmatrix}
\begin{bmatrix}
1&0&0\\
0&c_\phi&-s_\phi\\
0&s_\phi&c_\phi
\end{bmatrix}\\
&=\begin{bmatrix}
c_\theta c_\psi&s_\phi s_\theta c_\psi-c_\phi s_\psi&c_\phi s_\theta c_\psi+s_\phi s_\psi\\
c_\theta s_\psi&s_\phi s_\theta s_\psi+c_\phi s_\psi&c_\phi s_\theta s_\psi-s_\phi c_\psi\\
-s_\theta&s_\phi c_\theta&c_\phi c_\theta
\end{bmatrix}.
\end{aligned}
\end{equation*}
Here and below, we let $s_*=\sin(*)$, $c_*=\cos(*)$, and
$t_*=\tan(*)$. The final component of the state space is the angular
velocity vector $\Omega=[\Omega_1,\Omega_2,\Omega_3]^T$ of the
octorotor with respect to the body reference frame.

Next, we describe the control space of the octorotor. The control
space consists of the eight rotors, each of which exerts a force and
torque on the octorotor. For simplicity, we assume that the thrust of
each rotor can be directly set, and we do not model the motor
dynamics. Also, we assume that each rotor can rotate only in one
direction as shown in Figure \ref{Fig:octo_diag}, and that the thrust
exerted is in the upward direction $-b_3$ with respect to the
octorotor. Let $f_j$, $j=1,\ldots,8$, denote the forces exerted by the
rotors. We assume that each $f_j$ is restricted to an interval
$[f_{min},f_{max}]$, where $f_{min}$ and $f_{max}$ are the least and
greatest possible forces that each rotor can exert on the octorotor,
respectively. Since we are assuming that the rotors can only rotate in
one direction, we have $f_{min}\geq0$. Let $d$ denote the distance
from the center of mass of the octorotor to the center of each rotor, and
note that the thrust of the $j$th rotor induces a torque of magnitude
$d\cdot f_j$ about an axis in the octorotor plane perpendicular to the
arm containing the rotor. The $j$th rotor also induces a torque about
the $b_3$ axis of magnitude $c\cdot f_j$, where $c$ is the ratio of
induced torque to thrust. In particular, the rotors that rotate
counterclockwise induce a torque in the direction of $b_3$, and those
that rotate clockwise induce a torque in the direction of $-b_3$.

Let $F$ denote the total thrust exerted by the rotors, and let
$\tau_1,\tau_2,\tau_3$ denote the net torque exerted by the rotors
about the body axes $b_1,b_2,b_3$, respectively. Then letting
$\gamma=\frac{\pi}{8}$, the relationship between
$u=[F,\tau_1,\tau_2,\tau_3]^T$ and $f_{all}=[f_1,f_2,\ldots,f_8]^T$ is
given by $u=\Lambda f_{all}$, where $\Lambda$ is the matrix
\[{\footnotesize
\begin{bmatrix}
1&1&1&1&1&1&1&1\\
ds_\gamma&-ds_\gamma&-dc_\gamma&-dc_\gamma&-ds_\gamma&ds_\gamma&dc_\gamma&dc_\gamma\\
dc_\gamma&dc_\gamma&ds_\gamma&-ds_\gamma&-dc_\gamma&-dc_\gamma&-ds_\gamma&ds_\gamma\\
c&c&-c&-c&c&c&-c&-c
\end{bmatrix}}.\]

Next, let $m\in\R$ denote the mass of the octorotor,
$g=9.81\,\frac{\mathrm{m}}{\mathrm{s}^2}$ be the gravitational
  acceleration, and
\[J=\begin{bmatrix}
J_1&0&0\\
0&J_2&0\\
0&0&J_3
\end{bmatrix}\]
be the inertia matrix of the octorotor with respect to its body frame. Then let $\Delta_r=[\Delta_x,\Delta_y,\Delta_z]^T$ and $\Delta_R=[\Delta_{R,1},\Delta_{R,2},\Delta_{R,3}]^T$ denote unstructured force and torque disturbances due to dynamics that are unaccounted for. In particular, we view these disturbances as encompassing aerodynamic effects such as air drag and blade flapping, which, for simplicity, we do not model in detail. (See, e.g., \cite{HHWT, MKC} for more information.) We assume the disturbances are unknown, untracked, and measurable functions of time, and satisfy the inequalities $|\Delta_x|,|\Delta_y|,|\Delta_z|\leq\Delta_{r,max}$, and $|\Delta_{R,1}|,|\Delta_{R,2}|\leq\Delta_{R,12,max}$, $|\Delta_{R,3}|\leq\Delta_{R,3,max}$, where $\Delta_{r,max}$, $\Delta_{R,12,max}$, and $\Delta_{R,3,max}$ are fixed bounds. Then the differential equations describing the motion of the octorotor are
\begin{equation*}
\begin{aligned}
\dot{r}&=v\\
m\dot{v}&=\begin{bmatrix}
0\\ 0\\ mg
\end{bmatrix}
-F
\begin{bmatrix}
c_\phi s_\theta c_\psi+s_\phi s_\psi\\
c_\phi s_\theta s_\psi-s_\phi c_\psi\\
c_\phi c_\theta
\end{bmatrix}
+\Delta_r\\
\begin{bmatrix}
\dot{\phi}\\ \dot{\theta}\\ \dot{\psi}
\end{bmatrix}
&=\begin{bmatrix}
1& s_\phi t_\theta & c_\phi t_\theta \\
0 & c_\phi & -s_\phi \\ 
0 & \frac{s_\phi}{c_\theta} & \frac{c_\phi}{c_\theta}
\end{bmatrix}
\begin{bmatrix}
\Omega_1\\ \Omega_2\\ \Omega_3
\end{bmatrix}\\
J\dot{\Omega}&=\tau-\Omega\times J\Omega+\Delta_R.
\end{aligned}
\end{equation*}

In this paper, we focus on the dynamics of the components of the state vector
\[s=[v_z,\phi,\theta,\psi,\Omega_1,\Omega_2,\Omega_3]^T,\]
which we consider to be the ``inner loop'' components of the octorotor system. Note that the derivative of these components do not depend on $[x,y,z,v_x,v_y]$, and so we may safely view the components of $s$ as forming the state space of a well-defined control system.

\section{Controller}\label{Sec:ctrl}

\begin{figure}
\centering
\includegraphics[width=0.98\columnwidth]{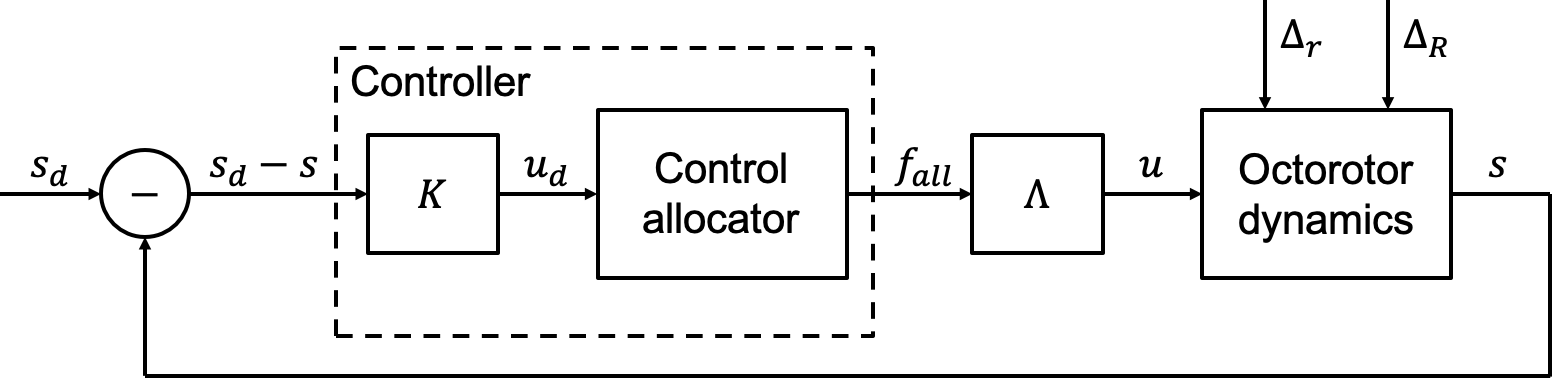}
\caption{Octorotor system block diagram}
\label{Fig:block_diagram}
\end{figure}

As mentioned above, we focus on controlling the dynamics of the octorotor's vertical velocity and orientation. In particular, the controller we use is designed to have the components $(v_z,\phi,\theta,\psi)$ of the octorotor state track the command $(v_{z,d},\phi_d,\theta_d,\psi_d)$. In other words, we wish for $s$ to track $s_d=[v_{z,d},\phi_d,\theta_d,\psi_d,0,0,0]^T$. The block diagram of the full control system is shown in Figure \ref{Fig:block_diagram}.

The controller consists of two components. The first, block $K$ in the
diagram, takes as input the difference $s_d-s$ between the current
state and the commanded state, and returns the commanded vector
$u_d=[F_d,\tau_{1,d},\tau_{2,d},\tau_{3,d}]^T$ of net force and
torques. This component disregards whether the octorotor is actually
capable of executing $u_d$, that is, whether there exists a valid
$f_{all}$ which yields the desired command. So $u_d$ is not
necessarily the actual control input
$u=[F,\tau_1,\tau_2,\tau_3]^T$. To find $u$, we include a second
component, the control allocator, which takes $u_d$ as input and
returns a value for $f_{all}$ such that $f_{min}\leq f_j\leq f_{max}$
for all $j$ and the resulting control input $u=\Lambda f_{all}$ is
intended to equal, or be close to, $u_d$. Note additionally that we
use the control allocator to implement rotor failures. Specifically,
we assume that the rotor failures are always known and the control
allocator is defined so that the thrusts $f_j$ corresponding to the
failed rotors match the values at which they are stuck. 

While under some circumstances the commanded value $u_d$ is not equal
to $u$, we want to ensure this never happens for the purposes of
verifying safety. In particular, we wish to find an invariant set for
which all the commanded values $u_d$ can be executed by the
octorotor. This allows us to split up our verification procedure into
$2$ separate steps as follows. The first step is to verify that a
chosen candidate invariant set is rendered invariant by the controller
under the assumption that $u$ is always equal to $u_d$. The second
step is to verify that the equality $u=u_d$ is in fact satisfied for
the control allocator for all states in the candidate invariant
set. This step is done for each rotor failure combination analyzed. 

\subsection{Controller block $K$}\label{Ssec:K}

The block $K$ controller component has a simple PD structure very
similar to \cite{MWY}, and is defined by the following:
\begin{equation}\label{ctrl}
\begin{aligned}
F_d&=\frac{mg}{c_\phi c_\theta}+K_{dz}(v_z-v_{z,d})\\
\tau_{1,d}&=-K_{p\phi}(\phi-\phi_d)-K_{d\phi}\Omega_1\\
\tau_{2,d}&=-K_{p\theta}(\theta-\theta_d)-K_{d\theta}\Omega_2\\
\tau_{3,d}&=-K_{p\psi}(\psi-\psi_d)-K_{d\psi}\Omega_3.
\end{aligned}
\end{equation}
Note that $mg/c_\phi c_\theta$ is the total rotor thrust needed to
keep the vertical acceleration of the octorotor at $0$. We take the
various coefficients $K_\cdot$ from a linear quadratic regulator that
we compute for the linearized octorotor dynamics at $s=0$, and under
the assumption that $v_{z,d},\phi_d,\theta_d,\psi_d,\Delta_{r,max}$,
$\Delta_{R,12,max},\Delta_{R,3,max}$ are all equal to $0$.

\subsection{Control allocator}\label{Ssec:allocation}

To find the individual rotor thrusts exerted by the octorotor in
response to the commanded control input $u_d$, we use a simple
pseudo-inverse control allocation method (see \cite{ODB}). At this
stage, we account for the possibility that some of the rotors have
failed. So let $W\subseteq\{1,\ldots,8\}$ be the (possibly empty) set
of indices corresponding to rotors that have failed, and for $j\in W$,
let $\bar{f}_j\in[f_{min},f_{max}]$ be thrust at which rotor $j$ is
stuck. Also let $\bar{f}_j=0$ for $j\notin W$ and
$\bar{f}_{all}=[\bar{f}_1,\ldots,\bar{f}_8]^T$. Next, define $\Lambda_W$
to be the matrix with the same entries as $\Lambda$, except that
column $j$ of $\Lambda_W$ is zeroed out for all $j\in W$. Then let
$\Lambda_W^\dagger$ be the pseudo-inverse of $\Lambda_W$. Now we
define $\tilde{f}_{all}=[\tilde{f}_1,\ldots,\tilde{f}_8]^T$ by 
\begin{equation}\label{tildefall}
\tilde{f}_{all}=\bar{f}_{all}+\Lambda_W^\dagger(u_d-\Lambda\bar{f}_{all}).
\end{equation}
The vector $\tilde{f}_{all}$ consists of potential rotor values, with
$\tilde{f}_j=\bar{f}_j$ for all $j\in W$ as required. However, it is
possible that $\tilde{f}_j$ falls outside $[f_{min},f_{max}]$ for some
$j\notin W$. We thus define $f_j$ for $j=1,\ldots,8$ by
\[f_j=\begin{cases}
\bar{f}_j & \text{if $j\in W$}\\
\midd\{f_{min},\tilde{f}_j,f_{max}\} & \text{if $j\notin W$,}
\end{cases}\]
where $\midd\{f_{min},\tilde{f}_j,f_{max}\}$ denotes the middle value
of the set $\{f_{min},\tilde{f}_j,f_{max}\}$. Thus
$f_j\in[f_{min},f_{max}]$ for all $j\notin W$, and so we let
$f_{all}=[f_1,\ldots,f_8]^T$. This completes the definition of the
control allocator.

To conclude this section, we wish to describe conditions for which
$u=u_d$. We first note that the matrix $\Lambda_W$ is of rank $4$ if
$|W|<4$, or $|W|=4$ and $W$ is not equal to $\{1,2,3,8\}$,
$\{1,2,4,7\}$, $\{1,2,5,6\}$, or any index set that can be obtained
from one of these sets by increasing each element by a common even
number and taking the remainder upon division by $8$ when
necessary. (This is equivalent to rotating the rotor failure patterns
corresponding to the $3$ sets about the octorotor center by multiples
of $90$ degrees.) When $\Lambda_W$ is of rank $4$, we have
$\Lambda_W\Lambda_W^\dagger=I_{4\times4}$ and
$\Lambda\tilde{f}_{all}=u_d$. So to determine whether $u=u_d$, it
suffices to check whether $\tilde{f}_j\in[f_{min},f_{max}]$ for all
$j\notin W$, and this is the condition we use, since we only consider
scenarios with up to $2$ rotor failures. 

\section{Barrier Functions}\label{Sec:barrier}

The safety conditions that we wish to prove rely on the notion of
invariant sets. An invariant set for a dynamical system is a region
$I$ of the state space of the system such that if the state begins in
$I$, then the state remains in $I$ at all future times. Our goal is to
ensure that for all commands $s_d$ one wishes to give the octorotor,
there exists an invariant set of octorotor states $s$ which are close
to $s_d$. Obtaining and verifying such invariant sets allows one to
gauge with certainty how close the controller is able to keep the
octorotor state to the given commands in spite of the disturbances
given in the dynamics model. 

To find these invariant sets, we utilize barrier functions. The type
of barrier function we use is a form of exponential barrier function
as introduced by \cite{NS}. (See, in particular, Remark 5, Proposition
1, and the related discussion.) With this concept, we find a set of
linear inequalities on the state space which forms a candidate
invariant region which we can verify. In this section, we provide a
brief overview of the reasoning behind exponential barrier functions,
and then explain the general form of exponential barrier functions we
use. 

Let $\dot{x}=f(x)$ define a dynamical system on some set
$X\subseteq\R^n$, and let $h=h_0:X\rar\R$ be a differentiable
function. The basic logic underlying exponential barrier functions is
that, for a trajectory $x(t)$ of the dynamical system, $h(x(t))\geq0$
for all $t\geq0$ whenever $h(x(0))\geq0$ and
\begin{equation*}
  \begin{aligned}
    &\frac{d}{dt}(h(x(t)))+p_1h(x(t))\\
    &=(\nabla h)(x(t))\cdot f(x(t))+p_1h(x(t))\geq0
  \end{aligned}
\end{equation*}
for all $t\geq0$, where $p_1>0$ is a constant. Thus, if one seeks an
invariant subset of $\{x\in X:h(x)\geq0\}$, one can, if necessary,
enforce the extra inequality $h_1(x):=(\nabla h)(x)\cdot
f(x)+p_1h(x)\geq0$, and form the set $\{x\in
X:h_0(x),h_1(x)\geq0\}$. If one can then prove that whenever the
initial condition is chosen from this set, the inequality
$h_1(x)\geq0$ holds indefinitely, then $h_0(x)\geq0$ holds
automatically. One can furthermore inductively form a sequence of
functions $h_j$ defined by $h_j(x)=(\nabla h_{j-1})(x)\cdot
f(x)+p_jh_{j-1}(x)$ and examine the sets $\{x\in
X:h_j(x)\geq0,j=0,\ldots,n\}$ to see if any of them are invariant, and
in particular, checking if the inequality $h_n(x)\geq0$ holds
indefinitely when $h_0(x),\ldots,h_n(x)\geq0$ holds at the initial
condition. See \cite{NS} for more details regarding the reasoning
above in the context of controlled dynamical systems. 

In light of the above discussion, we now explain the general barrier
function framework we use in constructing invariant sets for the
octorotor. It relies on a few starting assumptions. First, we assume
that a controller is already in place (namely, one of the form
described in Section \ref{Sec:ctrl}), which seeks to track a fixed
commanded state, and the dynamics are governed by a differential
equation of the form $\dot{x}=f(x,x_d,d)$, where $x\in\R^n$ is the
state, $x_d\in\R^d$ is the commanded state, $d\in\R^m$ denotes any
potential disturbances (all three being vertical vectors), and $f$ is
Lipschitz on any candidate invariant set of interest. Next, we assume
that the dynamics can be approximated sufficiently well by a
linearization of the form $\dot{x}=A(x-x_d)$, $A\in\R^{n\times n}$. In
the case of the octorotor, we assume $A$ results from the
linearization of the dynamics where $s=0$. Next, we assume that the
conditions we desire to be enforced are of the form
$P_i(x-x_d)+D_i\geq0$, where $P_i$ is a horizontal vector of length
$n$, and $D_i$ is a positive scalar. The aim of these inequalities is
to form a starting set of states $x$ nearby $x_d$ in which to find an
invariant set. The goal is then to confirm the ability of the
controller to keep the state $x$ near $x_d$. 

Now we form a sequence of barrier functions that aim to enforce
$P_i(x-x_d)+D_i\geq0$. First, we let $p_{i,j},\delta_{i,j}\geq0$ be
constants for $j=1,\ldots,n_i$ such that
$\sum_{j=1}^{n_i}\delta_{i,j}<D_i$. Then we define the sequence of
functions 
\begin{equation*}
\begin{aligned}
\tilde{h}_{i,0}(x)=\,&P_i(x-x_d)+D_i\\
\tilde{h}_{i,1}(x)=\,&P_i(I+p_{i,1}A)(x-x_d)+D_i-\delta_{i,1}\\
\tilde{h}_{i,2}(x)=\,&P_i(I+p_{i,1}A)(I+p_{i,2}A)(x-x_d)\\
\,&+D_i-\delta_{i,1}-\delta_{i,2}\\
\vdots\,&\\
\tilde{h}_{i,n_i}(x)=\,&P_i(I+p_{i,1}A)\cdots(I+p_{i,n_i}A)(x-x_d)\\
\,&+D_i-\delta_{i,1}-\cdots-\delta_{i,n_i}.
\end{aligned}
\end{equation*}
where $I$ is the $n\times n$ identity matrix. Notice that under the
linear dynamics,
$\tilde{h}_{i,j}(x)\leq\tilde{h}_{i,j-1}(x)+p_{i,j}\frac{d}{dt}(\tilde{h}_{i,j-1}(x))$
for $j=1,\ldots,n_i$, and so if the inequality $\tilde{h}_{i,j}\geq0$
holds, then $\tilde{h}_{i,j-1}$ holds automatically. Next, we let $\mu>0$
be a parameter and form our barrier function sequence 
\begin{equation*}
\begin{aligned}
h_{i,0,\mu}(x)&=\frac{P_i(x-x_d)}{D_i}+\mu\\
h_{i,1,\mu}(x)&=\frac{P_i(I+p_{i,1}A)(x-x_d)}{D_i-\delta_{i,1}}+\mu\\
h_{i,2,\mu}(x)&=\frac{P_i(I+p_{i,1}A)(I+p_{i,2}A)(x-x_d)}{D_i-\delta_{i,1}-\delta_{i,2}}+\mu\\
&\vdots\\
h_{i,n_i,\mu}(x)&=\frac{P_i(I+p_{i,1}A)\cdots(I+p_{i,n_i}A)(x-x_d)}{D_i-\delta_{i,1}-\cdots-\delta_{i,n_i}}+\mu,
\end{aligned}
\end{equation*}
and notice that $h_{i,j,1}$ is a postive scalar multiple of
$\tilde{h}_{i,j}$. We refer to the number $n_i$ as the depth of the
above sequence. Then $I(\mu)=\{x:h_{i,j,\mu}(x)\geq0,\forall i,\forall
j\}$ forms a parameterized family of candidate invariant sets, and
$I(1)$ in particular is the candidate set that enforces the conditions
$P_i(x-x_d)+D_i\geq0$. The purpose of the parameter $\mu$ is to allow
one to potentially establish a region of attraction for $I(1)$ by
showing that when the initial system state is in $I(\mu_{max})$ for
some $\mu_{max}>1$, the state will enter $I(\mu)$ for smaller and
smaller $\mu$ over time until it reaches $I(1)$. 

In order the confirm whether $I(\mu)$ is actually invariant for
$\mu\in[1,\mu_{max}]$, it is sufficient to check that for all $x\in
I(\mu_{max})$, we have that, for all $i$ and $j$, $h_{i,j,\mu}(x)=0$
implies that $\frac{d}{dt}(h_{i,j,\mu}(x))>0$. We shall refer to this
implication as the invariance condition for $h_{i,j,\mu}$. Under this
condition, it is clear that the least $\mu$ satisfying
$h_{i,j,\mu}(x)\geq0$ decreases with respect to time while
$\mu\geq1$.

One may notice that under the linearized dynamics, it is only
necessary to check the invariance condition for $h_{i,n_i,\mu}$ for
all $i$, since in this case, the fact that $h_{i,n_i,\mu}(x)\geq0$
holds implies that $h_{i,j,\mu}(x)\geq0$ holds for all $j\leq
n_i$. This continues to be true if we let $\delta_{i,j}=0$ for all $i$
and $j$. However, since we are interested in verifying invariance for
the original nonlinear dynamics, it is necessary to check the
invariance conditions for all functions. 
Also, under the nonlinear dynamics, the fact that
$h_{i,j,\mu}(x)\geq0$ holds may not automatically imply that
$h_{i,j-1,\mu}(x)\geq0$ for $\delta_{i,j}=0$, though this issue can be
mitigated by increasing $\delta_{i,j}$.

To conclude this section, we remark that in the case where $I(\mu)$ is
proven to be invariant for $\mu\in[1,\mu_{max}]$, it is straightforward
to check for an expanded region of attraction for $I(1)$ by verifying the
invariance conditions for $\mu\in[\mu_{max},\tilde{\mu}_{max}]$, where
$\tilde{\mu}_{max}>\mu_{max}$; and similarly one can check to see if
there are smaller invariant sets inside $I(1)$ by verifying the invariance
conditions for $\mu\in[\mu_{min},1]$, for $\mu_{min}\in(0,1)$. One can
then in a systematic way, e.g., through a bisection method,
approximate the exact $\mu$-interval over which the invariant conditions
are satisfied as accurately as desired; though SMT solving time may
limit the accuracy one can feasibly obtain.

\section{Octorotor Barrier Functions}\label{Sec:octo_barrier}

In this section, we describe in more detail the form of the barrier
functions we use for the octorotor system. As mentioned above, the
approximating linear dynamics we use is the linearization around
$s=0$. Conveniently, this approximation decomposes into $4$
subsystems. The first subsystem consists of only the vertical velocity
$v_z$. The nonlinear equation governing $v_z$, without disturbances,
is
\[\dot{v}_z=g-\frac{F}{m}c_\phi c_\theta,\]
which, when combined with the controller defined by \eqref{ctrl}, yields
\[\dot{v}_z=-\frac{K_{dz}c_\phi c_\theta}{m}(v_z-v_{z,d}).\]
(Throughout this section, we assume that $u=u_d$.) Hence, the linearized system for $v_z$ is
\[\dot{v}_z=-\frac{K_{dz}}{m}(v_z-v_{z,d}).\]
One of our goals is to ensure that $|v_z-v_{z,d}|\leq D_{v_z}$ for
some constant $D_{v_z}>0$, assuming the bound $|v_{z,d}|\leq
D_{v_{z,d}}$ with $D_{v_{z,d}}>0$. So we define our barrier function
components for $v_z$ by
\[h_{v_z,\mu}^\pm(s,v_{z,d})=\pm\frac{v_z-v_{z,d}}{D_{v_z}}+\mu,\]
one barrier function corresponding to $+$ and one to $-$. This means
that we do not make use of barrier functions of higher depth to
enforce $|v_z-v_{z,d}|\leq D_{v_z}$.

The next subsystem consists of the roll angle and rate
$(\phi,\Omega_1)$. The nonlinear equations governing these variables,
without disturbances, are
\[\dot{\phi}=\Omega_1+t_\theta(s_\phi\Omega_2+c_\phi\Omega_3),\quad
\dot{\Omega}_1=\frac{\tau_1+\Omega_2\Omega_3(J_2-J_3)}{J_1}.\]
Linearizing and combining with the controller \eqref{ctrl} yields
\[\dot{\phi}=\Omega_1,\quad
\dot{\Omega}_1=\frac{-K_{p\phi}(\phi-\phi_d)-K_{d\phi}\Omega_1}{J_1}.\]
Another goal is to ensure that $|\phi-\phi_d|\leq D_\phi$ for some
$D_\phi>0$, assuming that $|\phi_d|\leq D_{\phi_d}$, $D_{\phi_d}>0$,
and we aim to enforce the first inequality with barrier function
sequences of depth $1$. So for constants
$p_{\phi,1},\delta_{\phi,1}\geq0$ with $\delta_{\phi,1}<D_\phi$, we
define our barrier function components for $(\phi,\Omega_1)$ as the
following:
\begin{equation*}
\begin{aligned}
h_{\phi,0,\mu}^\pm(s,\phi_d)&=\pm\frac{\phi-\phi_d}{D_\phi}+\mu,\\
h_{\phi,1,\mu}^\pm(s,\phi_d)&=\pm\frac{\phi-\phi_d+p_{\phi,1}\Omega_1}{D_\phi-\delta_{\phi,1}}+\mu.
\end{aligned}
\end{equation*}

There are two remaining subsystems, one of which consists of the pitch
angle and rate $(\theta,\Omega_2)$, and the other consists of the yaw
angle and rate $(\psi,\Omega_3)$. Both subsystems are very similar to
that of the roll angle and rate. The nonlinear equations governing the
dynamics of these variables are
\begin{equation*}
\begin{aligned}
\dot{\theta}&=c_\phi\Omega_2-s_\phi\Omega_3, & \dot{\psi}&=\frac{s_\phi\Omega_2+c_\phi\Omega_3}{c_\theta}\\
\dot{\Omega}_2&=\frac{\tau_2+\Omega_1\Omega_3(J_3-J_1)}{J_2}, & \dot{\Omega}_3&=\frac{\tau_3+\Omega_1\Omega_2(J_1-J_2)}{J_3}.
\end{aligned}
\end{equation*}
Linearizing and then combining these equations with the controller \eqref{ctrl} yields
\begin{equation*}
\begin{aligned}
\dot{\theta}&=\Omega_2, & \dot{\Omega}_2&=\frac{-K_{p\theta}(\theta-\theta_d)-K_{d\theta}\Omega_2}{J_2}\\
\dot{\psi}&=\Omega_3, & \dot{\Omega}_3&=\frac{-K_{p\psi}(\psi-\psi_d)-K_{d\psi}\Omega_3}{J_3}.
\end{aligned}
\end{equation*}
Our goals for these variables is to ensure that $|\theta-\theta_d|\leq
D_\theta$ and $|\psi-\psi_d|\leq D_\psi$ for some $D_\theta,D_\psi>0$,
assuming that $|\theta_d|\leq D_{\theta_d}$ and $|\psi_d|\leq
D_{\psi_d}$, $D_{\theta_d},D_{\psi_d}>0$. As with the roll angle and
rate, we utilize barrier function sequences of depth 1. So for
constants
$p_{\theta,1},\delta_{\theta,1},p_{\psi,1},\delta_{\psi,1}\geq0$ with
$\delta_{\theta,1}<D_\theta$ and $\delta_{\psi,1}<D_\psi$, we define
the barrier function components for $(\theta,\Omega_2)$ and
$(\psi,\Omega_3)$ as the following:
\begin{equation*}
\begin{aligned}
h_{\theta,0,\mu}^\pm(s,\theta_d)&=\pm\frac{\theta-\theta_d}{D_\theta}+\mu,\\
h_{\theta,1,\mu}^\pm(s,\theta_d)&=\pm\frac{\theta-\theta_d+p_{\theta,1}\Omega_2}{D_\theta-\delta_{\theta,1}}+\mu\\
h_{\psi,0,\mu}^\pm(s,\psi_d)&=\pm\frac{\psi-\psi_d}{D_\psi}+\mu,\\
h_{\psi,1,\mu}^\pm(s,\psi_d)&=\pm\frac{\psi-\psi_d+p_{\psi,1}\Omega_3}{D_\psi-\delta_{\psi,1}}+\mu.
\end{aligned}
\end{equation*}
Lastly, we also include the following functions to ensure bounds of
the form $|\Omega_1|\leq D_{\Omega_1}$, $|\Omega_2|\leq D_{\Omega_2}$,
and $|\Omega_3|\leq D_{\Omega_3}$, where
$D_{\Omega_1},D_{\Omega_2},D_{\Omega_3}>0$:
\begin{equation*}
\begin{aligned}
h_{\Omega_1,\mu}^\pm(s)&=\pm\frac{\Omega_1}{D_{\Omega_1}}+\mu,\\
h_{\Omega_2,\mu}^\pm(s)&=\pm\frac{\Omega_2}{D_{\Omega_2}}+\mu,\\
h_{\Omega_3,\mu}^\pm(s)&=\pm\frac{\Omega_3}{D_{\Omega_3}}+\mu.
\end{aligned}
\end{equation*}
This completes the description of all the barrier functions.

\section{Verification process}\label{Sec:ver}

We will now formulate the statements about the octorotor system that
we formally analyze in dReal. The dReal SMT solver \cite{GKC} is
capable of analyzing statements in the theory of first-order real
arithmetic with non-linear function symbols (including, crucially,
transcendental functions) in order to determine their satisfiability.
In order to overcome well-known decidability issues with this theory,
dReal incorporates a numerical precision constant that mediates the
granularity of the proof search. Given a statement $\varphi$ in this
theory, dReal will reply with one of the following:
\begin{itemize}
\item An assignment of ranges (intervals) to the (free) variables of
  $\varphi$  that make the statement satisfiable modulo the numerical
  precision; or
\item A proof that there is no assignment that would make $\varphi$ satisfiable.
\end{itemize}
The first case is referred to as a \emph{$\delta$-SAT} result.  The second
case is referred to as an \emph{UNSAT} result. In general, it is not
possible to determine, given the assignment corresponding to a
$\delta$-SAT result whether or not there are values in the range of
the assignment that would make the statement satisfiable.  However,
one can sample from these intervals and evaluate the results in order
to further search for a genuinely satisfiable (\emph{SAT}) result.  In
practice, when running dReal we always carry out a na\"ive search
within $\delta$-SAT ranges by polling the midpoint of the returned
box.  Thus, in the sequel \emph{SAT} means that we have found a
genuine counter-example via this further polling of the variable
ranges returned as part of one of dReal's $\delta$-SATs.  We always
formulate safety properties in such a way that the goal is to obtain
an UNSAT result, i.e., a proof of unsatisfiability, from dReal.

First, we introduce a constant
$\epsilon$, a small positive number that we use for various purposes,
one of which is to make our safety conditions slightly more strict to
account for potential numerical errors. Next, we define a search space
over which the statements we formulate are proven. In particular, we
restrict all of the following variables to a symmetric interval about
$0$:\footnote{The regions could be asymmetric and for other kinds of
  tasks that might be more natural.  For the task at hand, symmetry of
  the regions is most reasonable.}
\[v_z,\phi,\theta,\psi,\Omega_1,\Omega_2,\Omega_3,\Delta_z,\Delta_{R,1},\Delta_{R,2},\Delta_{R,3},v_{z,d},\phi_d,\theta_d,\psi_d.\]
That is, for all variables $\xi$ above, we restrict $\xi$ to lie in
the interval $(-\xi_{max},\xi_{max})$, where $\xi_{max}>0$. We also
restrict the variable $\mu$ to the interval
$(1-\epsilon,\mu_{max}+\epsilon)$. This ensures that dReal conducts a
well bounded search, which avoids unnecessary computational
issues. As mentioned above, all conditions are expressed in terms of a
search for a state which \emph{violates} the desired
safety condition. The aim is then to use dReal to verify that such a
violation cannot exist, which proves the condition.

\subsection{Barrier function support is inside search space}
Now the first thing we wish to ensure is that the common support of
the barrier functions we define is a subset of our search space. This
is because it is necessary to search over the full
boundary of the support when checking invariance. For this property,
we formulate the dReal search as
follows. We search for a state $s$ near the boundary of the search
space ---i.e., satisfying $\xi>\xi_{max}-\epsilon$ or $\xi<-\xi_{max}+\epsilon$
for some $\xi\in\{v_z,\phi,\theta,\psi,\Omega_1,\Omega_2,\Omega_3\}$---
such that there exists a barrier function component $h$ and, if
needed, associated command $c$ such that $s$ is in or near the support
of $h(\cdot,c)$, i.e., $h(s,c)>-\epsilon$. If $h$ has no argument for
a command $c$, then the last inequality should be replaced by
$h(s)>-\epsilon$. If dReal verifies that no such state $s$ exists,
then we can conclude that the full support is inside the search
space.

\subsection{Invariance of barrier function support}
Next, we formulate the conditions for checking that the barrier
function support is invariant, and we do this by checking the
invariance property for each barrier function component. Here we are
assuming that the controller defined by \eqref{ctrl} always works,
i.e., $u=u_d$ and we disregard the rotor thrust limits. Also, we
assume that the states $s$ we search over are in or near the support
of every barrier function component, i.e., for all barrier function
components $h$, we have $h(s,c)>-\epsilon$ for some associated command
$c$, if such a command is needed, or otherwise $h(s)>-\epsilon$. With
these assumptions in mind, we formulate one condition for each barrier
function component. For the component $h_*$, we search for a state $s$
and, if needed, a command $c$ associated with $h_*$ such that $s$ is
on or near the boundary of the support of $h_*(\cdot,c)$ (or
$h_*(s)$), i.e., $-\epsilon<h_*(s,c)<\epsilon$ (or
$-\epsilon<h_*(x)<\epsilon$), and the time derivative of $h_*(s,c)$
(or $h_*(s)$) is almost negative, i.e.,
$\frac{d}{dt}(h_*(s,c))<\epsilon$ (or
$\frac{d}{dt}(h_*(s))<\epsilon$). For the sake of clarity, the
condition we impose on the state $s$ to consider it a violation of the
invariance condition for $h_*$ is the following: 
\begin{equation*}
\begin{aligned}
&\,\left({\text{for every b.f.\ component $h$, there exists an}\atop
\text{associated command $c$ such that $h(s,c)>-\epsilon$}}\right)\\
\wedge&\,\left({\text{there exists a command $c$ associated to $h_*$}\atop
\text{such that $h_*(s,c)<\epsilon$ and $\frac{d}{dt}(h_*(s,c))<\epsilon$}}\right).
\end{aligned}
\end{equation*}
For $h$ or $h_*$ that has no argument for a command in the above expression, we disregard the corresponding command $c$ and replace $h(s,c)$ with $h(s)$, or $h_*(s,c)$ with $h_*(s)$. If, for every barrier function component $h_*$, there is no state $s$ that satisfies the above condition, then the barrier function support is invariant under the controller defined by \eqref{ctrl}.

\subsection{Rotor thrust bounds}
The last condition we check is the equality $u=u_d$ we assumed in the previous condition. More specifically, we wish to check whether this equality is true over the support of the given barrier function and under the rotor failure combination being considered. As explained in Section \ref{Ssec:allocation}, it suffices to check whether the thrusts $\tilde{f}_j$ of the non-failed rotors given by the allocation method are valid and stay within the interval $[f_{min},f_{max}]$, that is, as long as $\rank(\Lambda_W)=4$. We formulate a condition for each combination of rotor failures, so let $W\subseteq\{1,\ldots,8\}$ such that $\rank(\Lambda_W)=4$. It is clear we can consider $u_d$ a function of $s$ by definition; and as a result, by \eqref{tildefall} we can view $\tilde{f}_{all}$ as a function of $s$. With this in mind, we can formulate our search as follows. We search for a state $s$ such that for every barrier function component $h$, $h(s,c)>-\epsilon$ for some corresponding command $c$ (or $h(s)>-\epsilon)$, and additionally, for some $j\notin W$, $\tilde{f}_j$ is outside, or nearly outside, of $[f_{min},f_{max}]$, i.e., $\tilde{f}_j-f_{max}>-\epsilon$ or $\tilde{f}_j-f_{min}<\epsilon$. If dReal verifies that no such state $s$ exists, then the octorotor controller behaves as intended for the rotor failure combination being analyzed.

\section{Results}\label{Sec:results}

We apply our verification process to an octorotor model with the characteristics of that in \cite{MWY}. Specifically, the octorotor model has mass $m=1.2\,\text{kg}$, inertia matrix entries $J_1=J_2=7.5\times10^{-3}\,\text{kg}\cdot\text{m}^2$ and $J_3=1.3\times10^{-2}\,\text{kg}\cdot\text{m}^2$, arm length $d=0.4\,\text{m}$, and torque-to-thrust ratio $c=(\frac{7.5}{3.13})\times10^{-2}\,\text{m}$. Additionally, we let the minimum and maximum rotor thrusts be $f_{min}=0$ and $f_{max}=\frac{1}{2}mg$, and let the maximum disturbance magnitudes be $\Delta_{r,max}=\frac{1}{10}mg$, $\Delta_{R,12,max}=(0.6\,\frac{\text{rad}}{\text{s}^2})J_1$, and $\Delta_{R,3,max}=(0.6\,\frac{\text{rad}}{\text{s}^2})J_3$.

Next, to find the coefficients of the controller \eqref{ctrl}, we compute the linear quadratic regulator for the linearization of the octorotor system at $s=0$, while assuming that $s_d=0$ and neglecting the disturbances $\Delta_r,\Delta_R$. Specifically, we find the linear controller that minimizes the objective function
\begin{equation*}
\begin{aligned}
&\int_0^\infty(s^TQs+|u|^2)\,dt,\\
&\text{with } Q=\diag\left(40,\frac{1}{4},\frac{1}{4},\frac{1}{4},\frac{1}{8},\frac{1}{8},\frac{1}{8}\right),
\end{aligned}
\end{equation*}
under the linearized dynamics. The resulting coefficients in \eqref{ctrl} are then (approximately) as follows:
\begin{equation}\label{K_coeffs}
\begin{aligned}
&K_{dz}=6.32,\quad K_{p\phi}=K_{p\theta}=K_{p\psi}=0.5,\\
&K_{d\phi}=K_{d\theta}=0.364,\quad K_{d\psi}=0.371.
\end{aligned}
\end{equation}

\begin{figure}
\centering
\includegraphics[width=0.9\columnwidth]{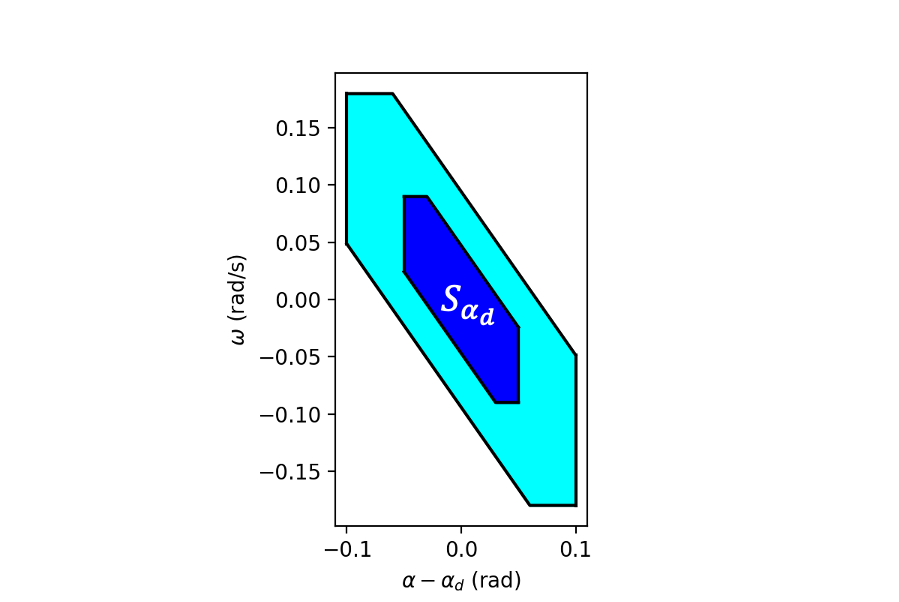}
\caption{The region $S_{\alpha_d}$}
\label{Fig:S_alpha_d}
\end{figure}

Next, to define the barrier function components, we make the following assignments:
\begin{equation*}
\begin{gathered}
D_{v_z}=0.25,\enspace D_{v_{z,d}}=1,\enspace D_\phi=D_\theta=D_\psi=0.05,\\
D_{\phi_d}=D_{\theta_d}=0.15,\enspace D_{\psi_d}=\epsilon,\enspace D_{\Omega_1}=D_{\Omega_2}=D_{\Omega_3}=0.09,\\
p_{\phi,1}=p_{\theta,1}=p_{\psi,1}=0.7,\enspace \delta_{\phi,1}=\delta_{\theta,1}=\delta_{\psi,1}=0.017,\\
\mu_{max}=2.
\end{gathered}
\end{equation*}
This means that the command $v_{z,d}$ is allowed to span the interval $(-1,1)$, and $\phi_d$ and $\theta_d$ are allowed to span $(-0.15,0.15)$. Note that there is no loss in letting $D_{\psi_d}=\epsilon$, hence $\psi_d\approx0$, since one can always redefine coordinates so that $\psi_d=0$. Also, for any combination of commands, the candidate invariant set resulting from the barrier functions is the set $I(1)$ of states $s$ such that $|v_z-v_{z,d}|\leq0.25$, $(\phi,\Omega_1)\in S_{\phi_d}$, $(\theta,\Omega_2)\in S_{\theta_d}$, and $(\psi,\Omega_3)\in S_{\psi_d}$, where
\[S_{\alpha_d}=\left\{(\alpha,\omega)\in\R^2:{|\alpha-\alpha_d|\leq0.05,|\omega|\leq0.09\atop |\alpha-\alpha_d+0.7\omega|\leq0.033}\right\}.\]
The set $S_{\alpha_d}$ is the dark blue region illustrated in Figure \ref{Fig:S_alpha_d}. The light blue region is the dilation of $S_{\alpha_d}$ by a factor of $2$, and corresponds to the region of attraction for the invariant set. Lastly, we let $\epsilon=10^{-8}$.

With all of the above definitions, we used dReal to obtain a proof of
safety in the case of no rotor failures, as well as many combinations
of rotor failures. That is, we used dReal to successfully verify that
the set $I(\mu)$ is invariant for $1\leq\mu\leq2$ under the controller
\eqref{ctrl} with coefficients \eqref{K_coeffs}, and that the
octorotor is capable of executing the controller's actions under
various rotor failure combinations. Table \ref{Tab:1} shows the times
it took to prove the various conditions that ensure safety in the case
where there are no rotor failures. Recall that UNSAT cases correspond
to safety proofs found and SAT cases correspond to concrete
counter-examples found.  (Note that there is a small
discrepancy between the sum of the times of the individual steps and
the total time since the latter is the run time for a single script
that performed all the steps in the table, and a small portion of the
script is not taken into account in the recorded times for the steps.)
Additionally, Table \ref{Tab:2} shows the various times it takes to
check whether the individual rotor thrusts $\tilde{f}_j$ stay in
$[f_{min},f_{max}]$ for various rotor failure combinations, in which
failed rotors exert zero thrust, i.e., they experience complete
failures. In particular, note that the octorotor remains safe for up
to two rotor failures, except when the failures occur on adjacent
rotors that rotate the same direction. Lastly, Table \ref{Tab:3} shows
the times it takes to check the individual rotor thrust bounds for
rotor failures where the thrusts of failed rotors can get stuck at 
zero or nonzero values. (Note that $\mu_{max}=2$ for all entries 
except where $\mu_{max}$ is specified, in which case $\mu_{max}$
was decreased in order to obtain an UNSAT.) All times were obtained
using $16$ cores.

\section{Conclusion}\label{Sec:con}

This paper introduces a framework for the formal verification of the
safety of control systems using exponential barrier functions, with
the target application of ensuring the faithful command tracking of an
octorotor controller. Our method uses barrier functions to construct
candidate invariant sets for the octorotor dynamics, and invariance of
these sets are then formally checked by the SMT solver dReal. We
account for potential rotor failures through a pseudo-inverse control
allocator, which we also verify to produce valid rotor thrusts via
dReal. Using our approach, we verify that a particular controller
causes the octorotor to follow commands within a certain margin of
error under dynamic disturbances and several types of rotor
failures. Our approach is fairly general, and can potentially be used
to construct safe invariant sets in many different systems and using
different kinds of controllers. One particularly promising application
is in the verification of gain scheduled controllers for aircraft.

\begin{table*}
\caption{Run times for checking safety conditions for no rotor failures}
\label{Tab:1}
\begin{center}
{\footnotesize
\begin{tabular}{ | l | l | l | l | l | l | }
\hline
\multirow{2}{*}{Step} & \multirow{2}{*}{Precision} & \multirow{2}{*}{SAT/UNSAT} & \multicolumn{3}{|c|}{Proof times (with 16 cores)} \\ \cline{4-6}
& & & real & user & sys \\ \hline
Search space contains b.f. support & $10^{-2}$ & UNSAT & 0m0.015s & \multicolumn{1}{|c|}{-} & \multicolumn{1}{|c|}{-} \\ \hline
Invariance, $v_z$ b.f. components & $10^{-2}$ & UNSAT & 0m0.165s & \multicolumn{1}{|c|}{-} & \multicolumn{1}{|c|}{-} \\ \hline
Invariance, $\phi,\Omega_1$ b.f. components & $10^{-2}$ & UNSAT & 253m20.942s & \multicolumn{1}{|c|}{-} & \multicolumn{1}{|c|}{-} \\ \hline
Invariance, $\theta,\Omega_2$ b.f. components & $10^{-2}$ & UNSAT & 253m41.323s & \multicolumn{1}{|c|}{-} & \multicolumn{1}{|c|}{-} \\ \hline
Invariance, $\psi,\Omega_3$ b.f. components & $10^{-2}$ & UNSAT & 0m0.055s & \multicolumn{1}{|c|}{-} & \multicolumn{1}{|c|}{-} \\ \hline
Rotor bounds (no rotor failures) & $10^{-2}$ & UNSAT & 0m0.109s & \multicolumn{1}{|c|}{-} & \multicolumn{1}{|c|}{-} \\ \hline
Total time & & & 507m3.938s & 7576m43.915s & 443m29.505s \\ \hline
\end{tabular}}
\end{center}
\end{table*}

\begin{table*}
\caption{Run times for checking rotor bounds with complete rotor failures}
\label{Tab:2}
\begin{center}
{\footnotesize
\begin{tabular}{ | l | l | l | l | l | l | }
\hline
\multirow{2}{*}{Step: Rotor bounds under failures} & \multirow{2}{*}{Precision} & \multirow{2}{*}{SAT/UNSAT} & \multicolumn{3}{|c|}{Proof times (with 16 cores)} \\ \cline{4-6}
& & & real & user & sys \\ \hline
Rotor $1$ complete failure & $10^{-2}$ & UNSAT & 0m0.925s & 0m5.847s & 0m4.571s \\ \hline
Rotor $1$, $2$ complete failures & $10^{-5}$  & SAT & 0m0.818s & 0m4.810s & 0m4.687s \\ \hline
Rotor $1$, $3$ complete failures  & $10^{-2}$ & UNSAT & 0m0.923s & 0m5.823s & 0m4.594s \\ \hline
Rotor $1$, $4$ complete failures  & $10^{-2}$ & UNSAT & 0m0.910s & 0m5.870s & 0m4.416s \\ \hline
Rotor $1$, $5$ complete failures  & $10^{-2}$ & UNSAT & 0m0.803s & 0m4.179s & 0m3.971s \\ \hline
Rotor $1$, $6$ complete failures  & $10^{-2}$ & UNSAT & 0m0.898s & 0m5.660s & 0m4.343s \\ \hline
Rotor $1$, $7$ complete failures  & $10^{-2}$ & UNSAT & 0m0.893s & 0m5.723s & 0m4.313s \\ \hline
Rotor $1$, $8$ complete failures & $10^{-2}$ & UNSAT & 0m1.035s & 0m6.969s & 0m4.322s \\ \hline
\end{tabular}}
\end{center}
\end{table*}

\begin{table*}
\caption{Run times for checking rotor bounds with stuck rotor failures}
\label{Tab:3}
\begin{center}
{\footnotesize
\begin{tabular}{ | l | l | l | l | l | l | }
\hline
\multirow{2}{*}{Step: Rotor bounds under failures} & \multirow{2}{*}{Precision} & \multirow{2}{*}{SAT/UNSAT} & \multicolumn{3}{|c|}{Proof times (with 16 cores)} \\ \cline{4-6}
& & & real & user & sys \\ \hline
\centering{Rotor $1$ thrust stuck at $\frac{mg}{8}$} & $10^{-2}$ & UNSAT & 0m0.946s & 0m5.245s & 0m4.499s \\ \hline
Rotor $1$ thrust stuck at $\frac{mg}{6}$, $\mu_{max}=1.6$ & $10^{-2}$  & UNSAT & 0m2.267s & 0m25.905s & 0m4.608s \\ \hline
Rotor $1$, $2$ thrusts stuck at $0$, $\frac{mg}{8}$ & $10^{-2}$ & SAT & 0m0.932s & 0m3.660s & 0m3.644s \\ \hline
Rotor $1$, $2$ thrusts stuck at $\frac{mg}{8}$, $\frac{mg}{8}$, $\mu_{max}=1.5$ & $10^{-2}$ & UNSAT & 0m7.261s & 1m45.881s & 0m5.297s \\ \hline
Rotor $1$, $2$ thrusts stuck at $0$, $\frac{mg}{6}$  & $10^{-5}$ & SAT & 0m0.887s & 0m5.530s & 0m4.155s \\ \hline
Rotor $1$, $2$ thrusts stuck at $\frac{mg}{6}$, $\frac{mg}{6}$  & $10^{-4}$ & SAT & 0m0.883s & 0m3.833s & 0m3.820s \\ \hline
Rotor $1$, $3$ thrusts stuck at $0$, $\frac{mg}{8}$, $\mu_{max}=1.6$ & $10^{-2}$ & UNSAT & 0m10.498s & 2m37.854s & 0m4.288s \\ \hline
Rotor $1$, $3$ thrusts stuck at $\frac{mg}{8}$, $\frac{mg}{8}$  & $10^{-2}$ & UNSAT & 0m1.270s & 0m9.844s & 0m3.996s \\ \hline
Rotor $1$, $3$ thrusts stuck at $0$, $\frac{mg}{6}$  & $10^{-6}$ & SAT & 0m0.863s & 0m3.947s & 0m4.064s \\ \hline
Rotor $1$, $3$ thrusts stuck at $\frac{mg}{6}$, $\frac{mg}{6}$, $\mu_{max}=1.3$ & $10^{-2}$ & UNSAT & 0m22.850s & 5m51.932s & 0m6.821s \\ \hline
Rotor $1$, $8$ thrusts stuck at $0$, $\frac{mg}{8}$  & $10^{-2}$ & UNSAT & 0m1.037s & 0m6.810s & 0m4.310s \\ \hline
Rotor $1$, $8$ thrusts stuck at $\frac{mg}{8}$, $\frac{mg}{8}$  & $10^{-2}$ & UNSAT & 0m3.899s & 0m51.320s & 0m4.599s \\ \hline
Rotor $1$, $8$ thrusts stuck at $0$, $\frac{mg}{6}$  & $10^{-2}$ & UNSAT & 0m1.502s & 0m13.834s & 0m4.068s \\ \hline
Rotor $1$, $8$ thrusts stuck at $\frac{mg}{6}$, $\frac{mg}{6}$, $\mu_{max}=1.1$  & $10^{-2}$ & UNSAT & 0m8.274s & 2m0.787s & 0m5.040s \\ \hline
\end{tabular}}
\end{center}
\end{table*}

\bibliographystyle{plain}
\bibliography{octorotorbib}

\begin{thebibliography}{10}

\bibitem{AE}
H.~{Alwi} and C.~{Edwards}.
\newblock Fault tolerant control of an octorotor using {LPV} based sliding mode
  control allocation.
\newblock In {\em 2013 American Control Conference}, pages 6505--6510, 2013.

\bibitem{AE2}
H.~{Alwi} and C.~{Edwards}.
\newblock Sliding mode fault-tolerant control of an octorotor using linear
  parameter varying-based schemes.
\newblock {\em IET Control Theory and Applications}, 9(4):618--636, 2015.

\bibitem{ACENST}
A.~D. {Ames}, S.~{Coogan}, M.~{Egerstedt}, G.~{Notomista}, K.~{Sreenath}, and
  P.~{Tabuada}.
\newblock Control barrier functions: Theory and applications.
\newblock In {\em 2019 18th European Control Conference (ECC)}, pages
  3420--3431, 2019.

\bibitem{AXGT}
A.~D. {Ames}, X.~{Xu}, J.~W. {Grizzle}, and P.~{Tabuada}.
\newblock Control barrier function based quadratic programs for safety critical
  systems.
\newblock {\em IEEE Transactions on Automatic Control}, 62(8):3861--3876, 2017.

\bibitem{FLL}
A.~{Freddi}, A.~{Lanzon}, and S.~{Longhi}.
\newblock A feedback linearization approach to fault tolerance in quadrotor
  vehicles.
\newblock In {\em Proceedings of the 18th IFAC World Congress}, pages
  5413--5418, 2011.

\bibitem{GKC}
Sicun Gao, Soonho Kong, and Edmund~M Clarke.
\newblock dreal: An smt solver for nonlinear theories over the reals.
\newblock In {\em International conference on automated deduction}, pages
  208--214. Springer, 2013.

\bibitem{GLL}
F.~{Goodarzi}, D.~{Lee}, and T.~{Lee}.
\newblock Geometric nonlinear {PID} control of a quadrotor {UAV} on {SE}(3).
\newblock In {\em 2013 European Control Conference (ECC)}, pages 3845--3850,
  2013.

\bibitem{HHWT}
G.~{Hoffmann}, H.~{Huang}, S.~{Waslander}, and C.~{Tomlin}.
\newblock Quadrotor helicopter flight dynamics and control: Theory and
  experiment.
\newblock In {\em Proceedings of the AIAA Guidance, Navigation, and Control
  Conference}, 2007.

\bibitem{JV2018}
O.~A. {Jasim} and S.~M. {Veres}.
\newblock Formal verification of quadcopter flight envelop using theorem
  prover.
\newblock In {\em 2018 IEEE Conference on Control Technology and Applications
  (CCTA)}, pages 1502--1507, 2018.

\bibitem{JV2019}
O.~A. {Jasim} and S.~M. {Veres}.
\newblock Nonlinear attitude control design and verification for a safe flight
  of a small-scale unmanned helicopter.
\newblock In {\em 2019 6th International Conference on Control, Decision and
  Information Technologies (CoDIT)}, pages 1652--1657, 2019.

\bibitem{KZC}
M.~{Khan}, M.~{Zafar}, and A.~{Chatterjee}.
\newblock Barrier functions in cascaded controller: Safe quadrotor control.
\newblock In {\em 2020 American Control Conference (ACC)}, pages 1737--1742,
  2020.

\bibitem{LLM}
T.~{Lee}, M.~{Leok}, and N.~H. {McClamroch}.
\newblock Nonlinear robust tracking control of a quadrotor {UAV} on {SE}(3).
\newblock {\em Asian Journal of Control}, 15(2):391--408, 2013.

\bibitem{MKC}
R.~{Mahony}, V.~{Kumar}, and P.~{Corke}.
\newblock Multirotor aerial vehicles: Modeling, estimation, and control of
  quadrotor.
\newblock {\em IEEE Robotics Automation Magazine}, 19(3):20--32, 2012.

\bibitem{MWY}
A.~{Marks}, J.~F. {Whidborne}, and I.~{Yamamoto}.
\newblock Control allocation for fault tolerant control of a {VTOL} octorotor.
\newblock In {\em Proceedings of 2012 UKACC International Conference on
  Control}, pages 357--362, 2012.

\bibitem{NHGAS}
Q.~{Nguyen}, A.~{Hereid}, J.~W. {Grizzle}, A.~D. {Ames}, and K.~{Sreenath}.
\newblock {3D} dynamic walking on stepping stones with control barrier
  functions.
\newblock In {\em 2016 IEEE 55th Conference on Decision and Control (CDC)},
  pages 827--834, 2016.

\bibitem{NS2}
Q.~{Nguyen} and K.~{Sreenath}.
\newblock Safety-critical control for dynamical bipedal walking with precise
  footstep placement.
\newblock In {\em IFAC Analysis and Design of Hybrid Systems}, pages 147--154,
  2015.

\bibitem{NS}
Q.~{Nguyen} and K.~{Sreenath}.
\newblock Exponential control barrier functions for enforcing high
  relative-degree safety-critical constraints.
\newblock In {\em 2016 American Control Conference (ACC)}, pages 322--328,
  2016.

\bibitem{NA}
P.~{Nilsson} and A.~D. {Ames}.
\newblock Barrier functions: Bridging the gap between planning from
  specifications and safety-critical control.
\newblock In {\em 2018 IEEE Conference on Decision and Control (CDC)}, pages
  765--772, 2018.

\bibitem{ODB}
M.~W. {Oppenheimer}, D.~B. {Doman}, and M.~A. {Bolender}.
\newblock Control allocation for over-actuated systems.
\newblock In {\em 2006 14th Mediterranean Conference on Control and
  Automation}, 2006.

\bibitem{RK}
M.~{Ranjbaran} and K.~{Khorasani}.
\newblock Fault recovery of an under-actuated quadrotor aerial vehicle.
\newblock In {\em 49th IEEE Conference on Decision and Control (CDC)}, pages
  4385--4392, 2010.

\bibitem{SLFFSS}
M.~{Saied}, B.~{Lussier}, I.~{Fantoni}, C.~{Francis}, H.~{Shraim}, and
  G.~{Sanahuja}.
\newblock Fault diagnosis and fault-tolerant control strategy for rotor failure
  in an octorotor.
\newblock In {\em 2015 IEEE International Conference on Robotics and Automation
  (ICRA)}, pages 5266--5271, 2015.

\bibitem{SLFSF}
M.~{Saied}, B.~{Lussier}, I.~{Fantoni}, H.~{Shraim}, and C.~{Francis}.
\newblock Fault diagnosis and fault-tolerant control of an octorotor {UAV}
  using motors speeds measurements.
\newblock {\em 20th IFAC World Congress}, pages 5263--5268, 2017.

\bibitem{SMGZ}
F.~{Sharifi}, M.~{Mirzaei}, B.~W. {Gordon}, and Y.~{Zhang}.
\newblock Fault tolerant control of a quadrotor uav using sliding mode control.
\newblock In {\em 2010 Conference on Control and Fault-Tolerant Systems
  (SysTol)}, pages 239--244, 2010.

\bibitem{WAE}
L.~{Wang}, A.~D. {Ames}, and M.~{Egerstedt}.
\newblock Safe certificate-based maneuvers for teams of quadrotors using
  differential flatness.
\newblock In {\em 2017 IEEE International Conference on Robotics and Automation
  (ICRA)}, pages 3293--3298, 2017.

\bibitem{WAE2}
L.~{Wang}, A.~D. {Ames}, and M.~{Egerstedt}.
\newblock Safety barrier certificates for collisions-free multirobot systems.
\newblock {\em IEEE Transactions on Robotics}, 33(3):661--674, 2017.

\bibitem{WTE}
L.~{Wang}, E.~A. {Theodorou}, and M.~{Egerstedt}.
\newblock Safe learning of quadrotor dynamics using barrier certificates.
\newblock In {\em 2018 IEEE International Conference on Robotics and Automation
  (ICRA)}, pages 2460--2465, 2018.

\bibitem{WS}
G.~{Wu} and K.~{Sreenath}.
\newblock Safety-critical control of a {3D} quadrotor with range-limited
  sensing.
\newblock In {\em ASME 2016 Dynamic Systems and Control Conference}, 2016.

\bibitem{XS}
B.~{Xu} and K.~{Sreenath}.
\newblock Safe teleoperation of dynamic {UAV}s through control barrier
  functions.
\newblock In {\em 2018 IEEE International Conference on Robotics and Automation
  (ICRA)}, pages 7848--7855, 2018.

\bibitem{XGTA}
X.~{Xu}, J.~W. {Grizzle}, P.~{Tabuada}, and A.~D. {Ames}.
\newblock Correctness guarantees for the composition of lane keeping and
  adaptive cruise control.
\newblock {\em IEEE Transactions on Automation Science and Engineering},
  15(3):1216--1229, 2018.

\bibitem{XWPGETGA}
X.~{Xu}, T.~{Waters}, D.~{Pickem}, P.~{Glotfelter}, M.~{Egerstedt},
  P.~{Tabuada}, J.~W. {Grizzle}, and A.~D. {Ames}.
\newblock Realizing simultaneous lane keeping and adaptive speed regulation on
  accessible mobile robot testbeds.
\newblock In {\em 2017 IEEE Conference on Control Technology and Applications
  (CCTA)}, pages 1769--1775, 2017.

\bibitem{ZC}
Y.~{Zhang} and A.~{Chamseddine}.
\newblock Fault tolerant flight control techniques with application to a
  quadrotor {UAV} testbed.
\newblock In T.~Lombaerts, editor, {\em Automatic Flight Control Systems -
  Latest Developments}, pages 119--150. InTech, 2012.

\bibitem{ZZRT}
Q.~{Zhou}, Y.~{Zhang}, C.~{Rabbath}, and D.~{Theilliol}.
\newblock Design of feedback linearization control and reconfigurable control
  allocation with application to a quadrotor {UAV}.
\newblock In {\em 2010 Conference on Control and Fault-Tolerant Systems
  (SysTol)}, pages 371--376, 2010.

\end{thebibliography}

\end{document}